\documentstyle[twocolumn,epsf,aps]{revtex}
\begin{document}
\title{Criteria of unconditional entanglement purification}
\author{Xiang-Bin Wang\thanks{email: wang$@$qci.jst.go.jp} 
\\
        Imai Quantum Computation and Information project, ERATO, Japan Sci. and Tech. Agency\\
Daini Hongo White Bldg. 201, 5-28-3, Hongo, Bunkyo, Tokyo 113-0033, Japan}

\maketitle 
\begin{abstract}
We show that the fidelity result of entanglement 
purification protocol
 with 2-way classical communications (C. H. Bennett et al, Phys. Rev. Lett., 76, 722(1996)) is always correct with whatever initial states.
We then give a general criteria on the unconditional entanglement 
purification: if a protocol satisfies our condition,
the fidelity result of the protocol with  the  product form of initial
state for the raw pairs is also correct for the same protocol with 
arbitrary initial state for of the raw pairs. 
With this general condition, we conclude that all existing purification
protocols can work unconditionally. 
\end{abstract}
Quantum entanglement is one of the most important resources in quantum 
information processing\cite{nil}. There are purification protocols
on how to distill almost perfect entangled pairs (EPR pairs) from a larger 
number lower quality pairs, by only local operations and classical communications. 
There are different types of entanglement purification 
protocols\cite{BDSW,bennett2,ekert,shorpre,gl}. In the earliest protocols
\cite{BDSW,bennett2,ekert}, it was assumed that those raw pairs shared by
Alice and Bob are in a product state: there are no entanglement among those 
pairs. The idea of entanglement purification was then used to the 
very important problem of security proof of quantum
key distribution(QKD)\cite{BB}: if Alice and Bob can share almost
perfect EPR pairs through purification, they can share a secret key by 
measuring each pairs in the same basis on each side. This idea first appeared
in Ref\cite{ekert}, with the advantage distillation protocol\cite{bennett2}.
  However, since the purification protocol itself
assumes the product state for the initially shared raw pairs, it was concerned
in Ref\cite{ekert} that:``if Eve provides pairs which are entangled with each 
other'', then the remained pairs after distillation ``may not converge to
the pure state $|\phi^+\rangle\langle\phi^+|$''.
It was then shown by Lo and Chau\cite{qkd} that the result 
of hashing protocol given by
Bennett et al\cite{BDSW} is actually correct with whatever form
 of initial states
therefore a type of security proof for QKD was strictly set up. After that,
Shor and Preskill show that Lo-Chau's security proof 
can be reduced to the BB84 
protocol which contains no difficult techniques such as
quantum storage or collective measurements. (We shall call this type
of protocol as prepare-and-measure protocol.)
In the whole security proof of the BB84\cite{BB} protocol based on viewpoint of
entanglement purification, there are two important steps: (1), the 
entanglement purification protocol actually works unconditionally, with
whatever type of initial state; (2), the purification
can be classicalized therefore on can reduce it to a type of
prepare-and-measure QKD protocol, the standard BB84.
Latter, the second step was studied in a more general background and
a general condition on classicalization was given\cite{gl}. 
With such a generalization, one does not have to specifically construct
the classicalization step in a new protocol. Instead, one may directly use
the main theorem in Ref\cite{gl}. However, so far there is no general 
statement on step 1, the unconditional
entanglement purification. Say, given a purification
protocol with product form of initial state for the raw pairs, under which
condition the fidelity result of the final pairs is still correct given 
whatever initial state for the raw pairs. In this Letter, we study this
problem and shall give a general condition for the unconditional entanglement
purification. With our theorem, we find that all existing purification
protocols can work unconditionally.

For clarity, we shall consider the details of a special case first, 
the (modified) advantage
distillation\cite{bennett2,ekert}. This is a 2-pair error-rejection protocol 
with 2-way classical communications (2-CC).
We then state our general theorem and consider other entanglement purification protocols
\cite{BDSW,gl} with our theorem. Note that so far no one has shown 
explicitly that
the 2-CC protocol works unconditionally with whatever type of channel
noise, though the question to
 protocol with 1 way classical communications (1-CC)\cite{BDSW}  has been studied\cite{qkd}.  Our main
contribution here is to give a general statement of the condition for
unconditional entanglement purification. Since our proof is closely related to Lo-Chau's 
security proof\cite{qkd}, we regard our result as generalized effects of Lo-Chau's security proof.
    
Let's first state our main idea.
If the initial state is not a product form, classical statistics does 
not necessarily work. Therefore we should not blindly assume anything
of classical statistics, including the error estimation from samples. 
To make everything clear, we shall start from a silly protocol where
Alice and Bob take a dark Bell measurement (i.e. a Bell measurement
without reading the measurement result)
to each raw pairs. After this step, we can safely define the rates of
each Bell states for the raw pairs by classical statistics. 
We  regard state $|\phi^+\rangle$ as 
the perfect EPR state. 
{\it Therefore in the case that a dark Bell measurement
has been done to each raw pairs, the state for raw pairs is in a product form.
} In the estimation, we let Alice and Bob
do collective measurements $XX,YY,ZZ$ to test pairs. These collective measurements are just the parity measurement in different bases, e.g., the measurement
result of $XX$ is  0 or 1 if the bit values in $X$ basis 
in two sides of the measured pair are same or different. 
Since all these parity measurements commute, classical sampling theory works exactly in estimating  the rate of different
Bell states. 
After measuring the test pairs, they know
the distribution over 4 Bell states for the remained raw pairs by the 
classical sampling theory. 
Also, in our initial protocol, all operations commute and
classical statistics works throughout the whole protocol. 
The fidelity between the finally distilled $m$ pairs and the $m$ products of
perfect EPR state ($|\phi^+\rangle$) is given by classical
statistics. We then show that the initial dark Bell measurement to each raw pairs
can be deleted and all the collective measurements involved in the protocol
 can be replaced by local measurement on $X,Y,Z$ basis in each side with
the fidelity of the distilled $m$ pairs being unchanged. Eve's information
is limitted by the Holevo bound\cite{qkd} therefore if the fidelity of the distilled 
$m$ pairs is sufficiently close to 1, her amount of information to the final key
is negligible.

We denote $Z,X,Y$ as measurement in the bases of 
$\{|0\rangle, |1\rangle\},$
$\{|\pm\rangle =\frac{1}{\sqrt 2}(|0\rangle\pm|1\rangle)\},
$$\{|y\pm\rangle =\frac{1}{\sqrt 2}(|0\rangle\pm i|1\rangle)\}$ respectively.
We use the notation $|\phi^{\pm}\rangle,|\psi^{\pm}\rangle$ for 4 Bell states 
$\frac{1}{\sqrt 2}(|00\rangle\pm|11\rangle),\frac{1}{\sqrt 2}(|01\rangle\pm|10\rangle)$. 
We shall use $WW$ for collective measurement to a pair in $W$ basis and $W\otimes W$ for local measurement
in $W$ basis in each side, where $W$
can be $X,Y$ or $Z$.
 In the purification protocol we assume intially Alice and Bob share 
$N$ raw pairs and our motivation is to distill $m$ very good pairs finally,
with its fidelity to state $\left(|\phi^+\rangle\langle \phi^+ 
|\right)^{m}$ being
exponentially close to 1. 
          
Consider the following (silly) entanglement purification protocol with everything clearly defined:
\\ {\it Protocol 1:}\\
Step 1: Entanglement distribution:
Alice prepares $N$ perfect EPR pairs, $|\phi^+;\rangle$. She sends half of each pair to Bob (over a noisy channey in general).
$N$ is a very large natural 
number.
Step 2: Dark Bell measurement: They take a Bell measurement to each pair but don't read the
measurement results.
Step 3: Error test. They randomly pick out $3k$ pairs and equally 
divide them into 3 groups. Note that $k$ is large but $k<< N$.  
They take collective measurement of $XX$ to each pair in the first group, $YY$
to each pair in the second group and $ZZ$ to each pair in the third group.
They then send the pairs in group 1,2,3 to trash can $Tx,Ty,Tz$, respectively. 
 If  too many of the measurement results are 1 they abort the protocol; 
otherwise
they continue.
Step 4: Error rejection.
a.) Bit-flip error rejection. They randomly 
group the remained pairs outside the trash cans with each group containing 2 pairs.
To each 2-pair group, they
 collect the parity
information to one pair (the destination pair) by a controlled-NOT
\cite{nil} operation on each side (bi-CNOT) in $Z-$basis. 
They take a collective measurement of $ZZ$
to the destination pair.
If the result is 0, 
they put the destination pair to the trash can
$Tz$ and keep the other pair; if it is 1, they send the destination pair to trash can $Tz$ and discard the other pair permenantly.
b.) Phase-flip error rejection. They randomly 
group the remained  pairs outside the trash cans
with each group containing 2 pairs.
To each 2-pair group, they
 collect the $X-$basis parity
information to the destination pair by an $X-$basis bi-CNOT operation. 
They take a collective measurement of $XX$
to the destination pair.
If the result is 0, 
they send the destination pair to the trash can
$Tx$ and keep the other pair, if it is 1, they send the destination
to trash can $Tx$ and discard the other pair permenantly. 
They do operations in a.) and b.) iteratively until they believe that
the state of the remained pairs outside the trash cans is very close to
a product state  of $|\phi^+\rangle\langle \phi^+|$.   
We assume there are $m$ pairs remained outside the trash cans then.\\

We call this protocol as a silly
 protocol because it is not useful in
any real entanglement purification or QKD, due to the fact that Alice and Bob
have to come together to make the collective measurements. However, it is an effective 
mathematical tool to clearly show the security of the final useful protocol.\\
 We regard a pair as a good pair if it is in state $|\phi^+\rangle$;
a bad pair if it is in any one of the other 3 
Bell states. Note that a good pair
always has the parity 0 in what ever bases of $X,Y,Z$, while there is always
a certain basis where a bad pair has the parity 1 for sure.    
After Step 2 all quantum 
entanglement among different pairs are removed therefore classical statistics works
exactly with all measurement bases being commute. 
In particular,
after Step 3, the probability of obtaining more than
$\delta N$ 1s by measurement $WW$ on the remained pairs and
 finding more than $(\delta - \epsilon_0)k$ 1s by measurement $WW$ on 
a group of
$k$ pairs is less than $\exp[-\frac{1}{4}\epsilon_0^2N/(\delta-\delta^2)]$.
Given the joint information of parity value distribution in 3 bases, one may
calculate the rate of good pairs and each of the 3  types of bad pairs 
in remained $N-3k$ raw pairs. 
 We denote $q_I,q_x,q_y,q_z$ for the rate of $|\phi^+\rangle, |\psi^+\rangle,
|\psi^-\rangle,|\phi^-\rangle$.
The different  rates of each of the 4 Bell states after and before bit-flip error rejection are related by the following formulas\cite{chau}
\begin{equation}
  \left\{ \begin{array}{rcl} 
q_I' & = & \displaystyle\frac{p_{I}^2 +
   p_{z}^2}{(q_{I} + q_{z})^2 + (q_{x} + q_{y})^2} , \\ \\
   q_x' & = & \displaystyle\frac{q_{x}^2 + q_{y}^2}{(q_{I} + q_{z})^2 +
   (p_{x} + p_{y})^2} , \\ \\
   q_y' & = & \displaystyle\frac{2q_{x} q_{y}}{(q_{I} + q_{z})^2 + (q_{x} +
   q_{y})^2} , \\ \\
   q_z' & = & \displaystyle\frac{2q_{I} q_{z}}{(q_{I} + q_{z})^2 + (q_{x} +
   p_{y})^2} .
  \end{array}
  \right. \label{errorrate}
 \end{equation}
Here the left hand sides are the quantities after bit-flip rejection operation and the symbols in
right hand sides
of each equations are for the quantities before   bit-flip rejection operation. Switching $q_x$ and
$q_z$ we shall obtain the relations after and before phase-flip rejection operation.
This shows that part a.) of Step 4 reduces the rate
of bad pairs
 of $|\psi^\pm\rangle$; part b.) reduces  the rate
of bad pairs of $|\phi^-\rangle,|\psi^-\rangle$.
Also, it can be shown that, the rate of remained pairs after the Bit-flip
rejection is
\begin{eqnarray}
f=\frac{1}{2}\frac{1}{(q_{I} + q_{z})^2 + (q_{x} +
   p_{y})^2}.
\end{eqnarray}
 If the initial
error rates are small (say, $q_x=q_y=q_z=t_0$), after $g$ 
rounds of error rejection, the rate of remained pairs is
in the magnitude order of $4^{-g}$ while the new bad-pair rate is in the magnitude order of $4^g t_0^{2^g}$. If $t_0$ is
small, a few steps of repetition of error-rejection is sufficient. 
This has been demonstrated numerically by Deutsch
et al\cite{ekert}.   We conclude that with an appropriate number 
of iteration, the state of the finally remained $m$ pairs  $\rho_m$
must be exponentially
close to $m$ perfect EPR pairs, i.e. 
\begin{eqnarray}
<\Phi_m|\rho_m|\Phi_m>= 1-\epsilon
\end{eqnarray} and $\epsilon$ is a small number 
exponentially close to 0, 
 $|\Phi_m\rangle = |\phi^+\rangle^{\otimes m}$.

The order of Step 2 and Step 3 can  be exchanged since the collective measurements in Step 3 are just coarse-grained Bell measurements which commute with
the dark Bell measurements in Step 2. For example, measurement $ZZ$
  is simply a 
measurement on whether the pair belongs to subspace $\{ |\phi^\pm\rangle \}$
or subspace  $\{ |\psi^\pm \rangle \}$. 

 Consider the bi-CNOT in $Z$ basis. On each side
they take the unitary operation of 
$|0\rangle|0\rangle\rightarrow |0\rangle|0\rangle, 
|0\rangle|1\rangle\rightarrow |0\rangle|1\rangle,|1\rangle|0\rangle\rightarrow |1\rangle|1\rangle,
|1\rangle|1\rangle\rightarrow |1\rangle|0\rangle$, with the second state being the destination qubit.
 Consequently, if the second pair is 
destination pair, a $Z-$basis bi-CNOT operation is a 
permutation in Bell basis as the following
\begin{eqnarray}
|\chi_{i,j}\rangle|\chi_{i',j'}\rangle\longrightarrow |\chi_{i\oplus i',j}\rangle|\chi_{i',j'\oplus j}\rangle.
\end{eqnarray}
Here $i,j$ can be either 0 or 1, 
 $|\chi_{0,0}\rangle$,$|\chi_{1,0}\rangle$,$|\chi_{0,1}\rangle$,$|\chi_{1,1}\rangle$ are notations for $|\phi^+\rangle$,
$|\phi^-\rangle$, $|\psi^+\rangle$,$|\psi^-\rangle$, respectively and 
symbol $\oplus$ is the calculation of  summation mod 2.
This permutation operation and the dark Bell measurement commute. {\it Proof}: Given arbitrary $l-$pair state $\rho_l$, one can always
write in the form $\rho_l
=\sum_l| \langle b_l|\rho_l|b_l\rangle\cdot b_l\rangle\langle b_l | + \hat O_{off}$, where $\{|b_l\rangle\}$ is the Bell bases for $l$ pairs, $l$ is from 1 to
$4^l$, $\hat O_{off}$ contains all off-diagonal terms in Bell bases which have no contribution to the outcome of  Bell measurement. 
With the original order of the protocol, i.e., first take dark measurement and then take
bi-CNOTs, the outcome state is 
$\rho_f=\hat P \left(\sum_l| \langle b_l|\rho_l|b_l\rangle\cdot b_l\rangle\langle b_l |\right)$
 and $\hat P$ is the permutation given by bi-CNOTs. If we reverse the order,
after the bi-CNOTs, the state is changed to 
$\hat P \left(\sum_l| \langle b_l|\rho_l|b_l\rangle\cdot b_l\rangle\langle b_l |\right) +\hat P (\rho_{off})$. Note that with the permutation
$\hat P$ in Bell bases, the off-diagonal terms will be still
off-diagonal and the diagonal terms will be still diagonal. 
If we then take the Dark Bell measurement, only the diagonal
terms survive. This is to say, with the order being reversed,
 the outcome is also $\rho_f$.  This makes the proof.\\ 
Similarly, one can  show that the $X-$basis bi-CNOT  also commutes with the dark Bell measurement.
Therefore {\it all bi-CNOT operations commute with the Bell measurement in Step 2}.  
 Therefore all operations in Step 4 and Step 2 commute, consequently
we can exchange the order of Step 2 and Step 4, i.e.
Step 2 in protocol 1 can be delayed until the end of the protocol.
This dark Bell measurement
in the end of the protocol can be deleted because it does not
change the fidelity of the remained $m$ pairs.
Explicitly, given any state of $l$ pairs, $\rho_l$, after a dark
Bell measurement to each pair, the  state is changed
  to $\rho_l'=\sum_i \langle b_i| \rho_l'|b_i\rangle \cdot |b_i\rangle\langle b_i|$, with $\{|b_i\rangle\}$ being $l-$Bell basis and $i$ running from 1 to
$4^l$. The fidelity of $\rho_l'$ is
\begin{eqnarray}
\langle b_1| \rho_l' | b_1\rangle = 
\langle b_1|
\sum_i \langle b_i| \rho_l'|b_i\rangle \cdot |b_i\rangle\langle b_i|
 b_1\rangle = \langle b_1| \rho_l | b_1\rangle
\end{eqnarray}  
and $|b_1\rangle=|\phi^+\rangle^{\otimes l}$. 
This shows that, although the dark Bell measurement  
may change the state of remained $m$ pairs, it does not change the fidelity.
Due to the above arguments, we now give the following modified protocol for QKD:
\\{\it Protocol 2:}\\
Step 1: Entanglement distribution.
Step 2: Error test. 
Step 3: Error rejection. 
Step 4: Alice and Bob send all trash cans to Eve.
Step 5: They measure $Z\otimes Z$ to each of the remained $m$ pairs outside 
the trash cans and use the result as an $m-$bit final key.\\
Step 1-3 are same with those in Protocol 1. 
As it has been argued already,  after Step 3 in protocol 2,
the state of remained $m$ pairs is exponentially close to 1.
Therefore Eve's 
information to the final key is upper bounded by a negligibly small number
$\epsilon'$
\cite{qkd}, even Eve 
controls everything except for the distilled $m$ pairs, 
including all  pairs in trash cans and everything else 
outside Alice and Bob's labs.

This protocol is still useless for any actual entanglement purification
 because it requires the collective
measurements though the step of
dark  Bell measurement to each pair has now been removed. Now we show that 
 if we replace all collective measurements $WW$ by local
measurements $W\otimes W$, the final result of the protocol is unchanged.

For such a purpose, we first show that Step 4 can be replaced by Step 4':\\
{\it Step 4'}: 
Alice and Bob don't send any states in trash cans to Eve. Instead they
take local measurement of $W\otimes W$ to the pairs in trash cans
$Tw$, ($Tw$ can be $Tx$, $Ty$ or $Tz$.)  and announce the results. 

If we only want an unconditionally secure final key, we can use the following
simple argument:
A restrictive Eve cannot do the attack better than an un-restricted Eve.
Eve in protocol 2 is an un-restricted Eve who controls everything 
except for the distilled $m$ pairs. 
Consider a restriction requires Eve
to measure $W\otimes W$ to each pairs in trash cans $Tw$ and announce the
measurement results.
  Note that except for this, Eve's action to everything else under her control
is not restricted, e.g., the ancillas which
had been prepared in the most initial stage 
when Alice distributes EPR pairs to Bob.
With the above restriction, Eve's information to the final key
must be also bounded  by $\epsilon'$. Or in other words, if Eve's information
is bounded by $\epsilon'$ with $whatever$ attack, her information must
be also bounded by $\epsilon'$ with any $specific$ type of attack.
Replacing Step 4 with Step 4' in protocol 2 is equivalent to
add the above restriction to Eve
in
protocol 2.
This is to say, it will cause no harm to the security of final key 
if Alice and Bob keep all trash cans and  take 
local measurement $W\otimes W$ to pairs in trash cans
$Tw$ and announce the measurement outcome. 

We can also show that Step 4 can be replaced by Step 4' by the fidelity criteria, 
from the viewpoint of entanglement purification. Consider the case that
 Step 4 is replaced 
Step 4'.
We only need to argue that the local measurement in Step 4'
may cause no
changes to the fidelity of the $m$ remained pairs. 
Suppose the state of the total space immediately
after Step 3
is ${\bf \rho_{t}} = \rho_{m,c,e}$ and the subscript $m, c, e$ denote for 
the subspace of $m$ remained pairs, trash cans and environment, respectively.
The state of the remained $m$ pairs immediately after Step 3
is 
\begin{eqnarray}
\rho_m=tr_{c,e} {\bf \rho_t}.
\end{eqnarray}
The notation $tr_{c,e}$ means the partial trace calculation
in subspaces of trash cans and environment.
 In the protocol, if the error test is passed,  Alice and
Bob will $always$ accept the  remained $m$ pairs no matter what
the measurement outcome they may obtain in Step 4'. 
Therefore for the quantity of fidelity of the remained $m$ pairs,
we only care about the $averaged$ state for the $m$ pairs after
after Step 4',
 with the average being taken over all possible measurement results
in Step 4'. Mathematically, the averaged total state after Step 4' is
\begin{eqnarray}
{\bf \rho_t'}= \sum_L \langle L|{\bf \rho_t}|L\rangle |L\rangle\langle L|.
\label{rt} 
\end{eqnarray}
Here $|L\rangle$ is an eigenstate of the local measurement to those
qubits in trash cans in Step 4'.
Therefore the averaged state for the remained $m$ pairs after Step 4'  is
\begin{eqnarray}
\rho'_m=tr_{c,e} {\bf \rho_t'}=tr_{e} \sum_L <L|{\bf \rho_t}|L>=tr_{c,e}  \rho_t=\rho_m.
\label{int}
\end{eqnarray}
This shows that the averaged state for the remained $m$ pairs 
after Step 4' is just $\rho_m$, therefore Step 4'
does not change the fidelity of the remained $m$ pairs.
 
Now we show why all collective measurements $WW$ can simply be replaced by local measurements
$W\otimes W$.
With Step 4 being replaced by Step 4', each pairs in trash cans are measured two times, the first is $WW$, the second is $W\otimes W$.
They commute, therefore one can exchange
the order of them. More over, once we have 
measured $W\otimes W$, we do not need to measure $WW$, 
since the result of $WW$
has been explicitly determined already by the measurement $W\otimes W$ . 
Therefore all the collective 
measurements in protocol 2 can be replaced by local measurements in each side. 
 We obtain the following
protocol with all measurements being local: 
\\{\it Protocol 3:}\\
Step 1: Entanglement distribution.
Step 2: Error test. They randomly pick out $3k$ pairs and equally 
divide them into 3 groups.   
They take local measurement of $X\otimes X,Y\otimes Y,Z\otimes Z$ 
to each pair in group 1,2,3, respectively. 
They announce the measurement results and discard those measured qubits.
If too many of the results from two sides disagree, the abort the protocol; 
otherwise
they continue.
Step 3:a.) Bit-flip error rejection. They randomly 
group the remained pairs with each group containing 2 pairs.
To each 2-pair group, they
 collect the parity
information to the destination pair
 by a $Z-$basis bi-CNOT operation. 
They take local measurement  $Z\otimes Z$ 
to the destination pair and announce the measurement results from two sides.
If they are same, 
they discard the destination pair and keep the control pair;
 if they are different they discard both pairs. 
b.) Phase-flip error rejection. They randomly 
group the remained  pairs 
with each group containing 2 pairs.
To each 2-pair group, they
 collect the $X-$basis parity
information to the destination pair by a $X-$basis bi-CNOT operation. 
They measure $X\otimes X$ 
to the destination pair and announce the measurement results from two sides.
If they are same, 
they discard the destination pair and keep the control pair;
 if they are different they discard both pairs.
They do operations in a.) and b.) iteratively and finally
 there are $m$ pairs remained outside the trash cans.
Step 4: They measure $Z\otimes Z$ to each of the remained $m$ pairs
and use the result as an $m-$bit final key.

So far we have given a detailed and rigorous proof of the 
entanglement purification and security of QKD with 2-CC. 
There are two important issues in the reduction. The first issue
is the removal of the initial dark Bell measurement: this measurement
commute with bi-CNOTs and all collective measurements of $WW$, 
therefore it can be delayed and removed, with the fidelity of the 
purification
outcome unchanged. The second issue is, after removal of dark Bell 
measurement, any collective measurements $WW$ can be replaced by
local measurement of $W$ on each side, i.e., $W\otimes W$, since
$WW$ and $W\otimes W$ commute. Therefore we propose the following
general theorem:

{\bf Theorem}: {\it Suppose we have certain fidelity result for the 
final pairs through
a certain purification protocol with product form of state for the raw 
pairs. We must have the same fidelity result for the same protocol 
with whatever form of state for
the raw pairs, provided that the protocol satisfies the following conditions:
(1) All quantum operations needed are not more than bi-CNOTs and local
measurements on each side, i.e. $\{W\otimes W\}$; (2)The fidelity
result is unchanged if we replace any local measurement $W\otimes W$
by collective measurement $WW$; (3) All $WW$  
commute with dark Bell measurement. }
  
Together with the main theorem in Ref\cite{gl}, this theorem works 
effectively on the security proof of any purification-based QKD protocol.
We can just consider the conditional purification first:
the special case of product state for the raw pairs
and consider the fidelity result of purification. After that, we use our 
theorem and the main theorem in Ref\cite{gl} to reduce it to the unconditional
entanglement  purification and moreover, prepare-and-measure QKD protocol.

To reduce protocol 3 
to a prepare-and-measure protocol,
we only need to replace the Phase-flip rejection part in Step 3 with
3-pair phase error correction used in Ref\cite{gl}.
There are only bi-CNOTs and collective measurements $X\otimes X,Z\otimes Z$ 
involved\cite{gl}.According to our theorem, the purification must be also
unconditional. Therefore the unconditional 
security of
QKD\cite{gl,chau,wang} protocols with 2-CC is confirmed.
Similarly, the entanglement purification protocol with hashing\cite{BDSW}
and the one with quantum error correction code\cite{shorpre}
also only contains bi-CNOTs and local measurement $W\otimes W$ therefore
the fidelity result of the protocol is unconditionally true with whatever
initial state. 
 The issue of side information
of Shor-Preskill protocol has been studied by Hwang\cite{hwang}. \\  
{\bf Acknowledgement:} I thank Prof Imai H for support. I thank H. P Yuen for 
 comments.


\begin{thebibliography}{99}
\bibitem{nil}M. A. Nielsen and I. L. {\it Chuang, Quantum Computation and Quantum Information}, Cambridge University Press, 2000.

\bibitem{BDSW} C. H. Bennett, D. P. DiVincenzo, J. A. Smolin,
and W. K. Wootters, Phys. Rev. A54, 3824(1996).
\bibitem{bennett2}C. H. Bennett, G. Brassard, S. Popescu, B. Schmacher, J. Smolin, and
W. K. Wooters, Phys. Rev. Lett., 76, 722(1996).
\bibitem{ekert}
 D.~Deutsch, A.~Ekert, R.~Jozsa, C.~Macchiavello,
S.~Popescu, and A.~Sanpera, Phys.~Rev.~Lett., 77, 2818(1996);
  Erratum
Phys.~Rev.~Lett. {\bf 80}, 2022 (1998).
\bibitem{shorpre} P. W. Shor and J. Preskill, Phys. Rev. Lett., vol. 85,441(2000).
\bibitem{gl} D. Gottesman and H.-K. Lo, IEEE Transactions on
 Information Theory, 49, 457(2003).
\bibitem{qkd} H.-K.~Lo and H.~F.~Chau,  Science,
283, 2050(1999). 
\bibitem{BB}
C. H. Bennett and G. Brassard, 
{\em Proceedings of IEEE International Conference on Computers, 
Systems and Signal Processing, Bangalore, India, 1984},  (IEEE Press,
1984), pp. 175--179;
C.H. Bennett and G. Brassard,
IBM Technical Disclosure Bulletin {\bf 28}, 3153--3163 (1985). 
\bibitem{chau} H. F. Chau, Phys. Rev. A66, 060302(R) (2002).
\bibitem{wang} X. B. Wang, Phys. Rev. Lett., 92, 077902(2004).
\bibitem{hwang}W. Y. Hwang, quant-ph/0402032.
\end{thebibliography}
\end{document}